\documentclass[12pt, letterpaper, twoside]{article}
\pdfoutput=1

\usepackage{graphicx}
\usepackage[numbers,sort&compress]{natbib}

\title{Deep magma transport control on the size and evolution of explosive volcanic eruptions}
\author
{Simone Colucci,$^{1\ast}$ Paolo Papale,$^{1}$ \\
\\
\normalsize{$^{1}$ Istituto Nazionale di Geofisica e Vulcanologia, sezione di Pisa}\\
\normalsize{Via Cesare Battisti, 53, 56125, Pisa, Italy}\\
\\
\normalsize{$^\ast$Simone Colucci; e-mail:  simone.colucci@ingv.it.}
}

\begin{document}

\maketitle

\section*{Abstract}
Explosive eruptions are the surface manifestation of dynamics that involve transfer of magma from the underground regions of magma accumulation. Evidence of the involvement of compositionally different magmas from different reservoirs is continuously increasing to countless cases. Yet, models of eruption dynamics consider only the uppermost portion of the plumbing system, neglecting connections to deeper regions of magma storage. Here we show that the extent and efficiency of the interconnections between different magma storage regions largely control the size of the eruptions, their evolution, the causes of their termination, and ultimately their impact on the surrounding environment. Our numerical simulations first reproduce the magnitude-intensity relationship observed for explosive eruptions on Earth and explain the observed variable evolutions of eruption mass flow rates. Because deep magmatic interconnections are largely inaccessible to present-day imaging capabilities, our results imply a limit to eruption size forecasts based on observations and measurements during volcanic unrest.

\section*{Introduction}
Explosive eruptions are characterised by a variety of mass flow rate evolutions (\cite{pouget2013} and references therein) and erupted mass spanning several orders of magnitude \cite{pyle2015}. The activation of multiple reservoirs during individual eruptions is extensively documented  (e.g., \cite{araya2019} and references therein) and accepted as a general reference model \cite{edmonds2019}. 
Still, dynamic models of volcanic eruptions consider at most one single, progressively emptied magma reservoir \cite{bower1998,folch1998,marti2000,macedonio2005,folch2009,colucci2014}. Such models strictly associate larger magma chambers to larger eruption intensity and longer eruption duration, and invariably predict progressive chamber pressure and mass flow-rate decrease with time. On the contrary, real eruptions show virtually any mass flow-rate (or intensity) evolution with time, requiring therefore a different approach to explain such first order observations.

 By considering multiple magma chambers mutually interacting during volcanic eruptions, an entirely new world appears, explaining the observed eruption magnitudes and intensities and the large variability of mass flow rate evolutions. A new fundamental quantity emerges, represented by the efficiency of magma transfer across the magmatic reservoirs.  
 That quantity largely controls eruption evolution and the causes of its termination, and ultimately the impacts on the surrounding environment, with substantial implications for volcanic hazards.

 \section{Numerical simulations}

Figure 1 (left) shows the simulation setup. A deep, large reservoir hosting andesitic magma is connected through a vertical planar dyke to a shallow, smaller chamber hosting more chemically evolved dacite, connected to the surface through a cylindrical conduit. The input parameters (volume and depth of the chambers, conduit diameter, dyke width, volatile abundance in the two magmas) are varied in the range reported in the figure caption, reflecting more frequently reported values for explosive eruptions in a wide literature. Our computations refer to only the quasi-steady phases during which large volcanic columns develop and evolve, with or without partial or total column collapses and generation of pyroclastic flows \cite{neri1998}. Provided that the system as a whole evolves on a time scale sufficiently longer than the transit time of magma along the simulated domain, system evolution can be approximated as a series of discrete steady flow steps \cite{papale2001,macedonio2005,colucci2017}, with boundary conditions at each step depending on the previous dynamics and reflecting system evolution. The computations are terminated when the pressure somewhere in the system falls below lithostatic by a quantity exceeding the rock tensile strength \cite{bower1998,Gudmundsson2007,Kabele2017} leading to rock collapse and either eruption closure or transition to highly transient dynamics. When that happens at chamber level, caldera collapse is expected \cite{Gudmundsson2007,Kabele2017}.

The right panels in Figure 1 show one example of calculations (only a few times are reported). At each time step the flow variable distributions display well-known behaviours (e.g., \cite{papale2001}) reflecting a large gradient friction-dominated region leading to magma fragmentation, followed by an inertia-dominated region with lower gradients close to conduit exit. Lumped pressure in the shallow chamber corresponds to both dyke exit and conduit base pressure, whereas magma mixing in the chamber results in a jump of gas volume fraction and (not shown) magma density, and by continuity, magma ascent velocity.

Figure 2, obtained using the Dakota software \cite{adams2014} through a latin hypercube sampling within the range of employed system conditions, reports the two computed major quantities (defined in the Methods section) magnitude (a measure of the mass of discharged magma) and intensity (a measure of mass flow-rate
). For eruptions with intensity $<10.5$, a reported dominance of unsteady dynamics (e.g., \cite{miller1994, pistolesi2016, gudnason2018, pioli2008}) likely explains their absence in the simulated set. Apart from such low intensity eruptions, the observed magnitude-intensity relationship is well reproduced from relatively small, frequent M4 to globally impacting, destructive, rare M8 eruptions, suggesting that our simulations capture, on a first order, the major factors controlling the evolution of explosive volcanic eruptions.  
The colored symbols refer to cases for which the termination of the simulation is due to deep (red) or shallow (blue) chamber collapse (termination at volcanic conduit level is rare in our computations, and none for the deep dyke). Although with ample superposition, and in line with observations, deep chamber collapse is mostly associated to large magnitude eruptions, and characterises about all of the globally impacting M7+ eruptions.

\section{Control by deep interconnections}

The extent of magma chamber interconnections determines the transfer capability of magma across the reservoirs, exemplified in our simulations by the dyke width parameter. Dyke width is therefore a proxy for the effective mass flow rate across a complex system of interconnections possibly involving multiple dykes and intermediate storage zones \cite{edmonds2019}. Accordingly, parameterization of dyke width means parameterizing the overall interconnection efficiency. With small dyke width magma transfer across reservoirs is poorly efficient, thus the eruption discharge rate is not compensated by deep magma arrival (Fig. 3b,d, cold colours). As a consequence, shallow chamber pressure decays rapidly (Fig. 3a), while deep chamber pressure changes only slightly (Fig. 3c) reflecting minor magma withdrawal from depth. At such conditions eruption closure quickly follows from shallow chamber pressure drop after progressive decline of the eruption discharge rate.

With larger dyke width (10-12 m in Fig. 3), the increased efficiency of deep mass transfer sustains the eruption discharge rate (Fig. 3b, d, light blue/cyan curves), leading to longer sub-steady eruption phase (up to about 8 hours in the figure with nearly constant mass flow rate). Deep chamber pressure decrease becomes significant (Fig. 3c) due to increased magma withdrawal, eventually reaching the rock collapse threshold (starting from the cyan curve, or dyke width of 12 m). 

With further increase in dyke width, the mass flow rate along the dyke can overcome the one along the shallow conduit (hot-coloured lines in Fig. 3b,d), causing deep chamber depressurization (Fig. 3c) and shallow chamber pressurization (Fig. 3a). In turn, the latter causes the eruption discharge rate to increase (Fig. 3b). However, shallow pressure increase and deep pressure decrease rapidly lead to decreased of mass flow rate along the dyke (Fig. 3d), as the concomitant action of increasing downflow and decreasing upflow pressure driving dyke flow. Therefore, shallow chamber pressurization is self-buffered, as it induces changes in conduit and dyke flow ending up with less pressure increase. Accordingly, after initial shallow pressurization and intensity increase, the eruption evolves towards shallow pressure decrease (Fig. 3a) and about constant or slightly decreasing mass flow rate (Fig. 3b).

\section{Shallow versus deep controlled eruption regimes}

The dynamics illustrated above show that besides explaining the observed magnitude-intensity relationship, the interplay between magmatic reservoirs provides the physical framework explaining  observed variabilities in the evolution of eruption intensity, with mass flow-rates that can decrease, obscillate, remain constant, or increase depending on such an interplay. Non-linear relationships dominate the dynamics:
with sufficiently small dyke width, its increase produces longer eruption duration; conversely, with larger dyke width, its further increase results in shorter eruption duration. The longest durations correspond to near-balance of conduit and dyke mass flow rates (Fig. 3b,d). Eruption magnitude (reported in panel 3b) is also non-linearly related to dyke width: for small dyke width, its increase results in larger eruption magnitude, whereas for dyke widths sufficiently large to cause eruption termination by deep chamber collapse, the eruption magnitude becomes poorly or not dependent on further dyke width changes.

This situation depicts the existence of two regimes for explosive volcanic eruptions (Fig. 4): a shallow-dominated regime determined by low inter-chamber magma transfer efficiency and evolving to shallow chamber collapse, characterised by rapid shallow chamber pressure decrease, minor changes in the deep reservoir, and positive dependence of eruption magnitude and duration on deep magma transfer efficiency; and a deep-dominated regime determined by large inter-chamber magma transfer efficiency and evolving to deep chamber collapse, characterised by initial shallow pressure and eruption mass flow rate increase, significant pressure decrease in the deep reservoir, negative dependence of eruption duration and poor or no dependence of eruption magnitude on magma transfer efficiency.

\section{Implications for volcanic hazard}

We have shown that the efficiency of magma transfer between reservoirs, controlled by deep interconnections, plays a central role in determining the eruption magnitude and evolution, and ultimately the impacts on the surrounding environment. Unfortunately, we do not have yet any means to image with sufficient accuracy the complexity of volcanic plumbing systems.In fact, we know little or nothing about the deep system of interconnections extending down towards deeper magmatic reservoirs. Our results show that without a reliable and accurate estimate of the corresponding magma transfer efficiency, we may never be able to forecast the eruption magnitude, intensity and duration, or its evolution towards large caldera collapse, no matter the type, intensity and evolution of the observed signals. As a matter of fact, no robust, generally accepted method to forecast the size of an impending eruption has been produced to-date. Previous analysis \cite{papale2018} suggests that the highly non-linear nature of volcanic processes limits deterministic predictions of volcanic eruption size. This study shows that even an the deterministic approach employed here leads to conclude that finding confident relationships between unrest dynamics and size of the impending eruption may continue to be a very hard task, if not a hopeless undertaking.

\clearpage

\begin{figure}
\center
\includegraphics[width=1\linewidth]{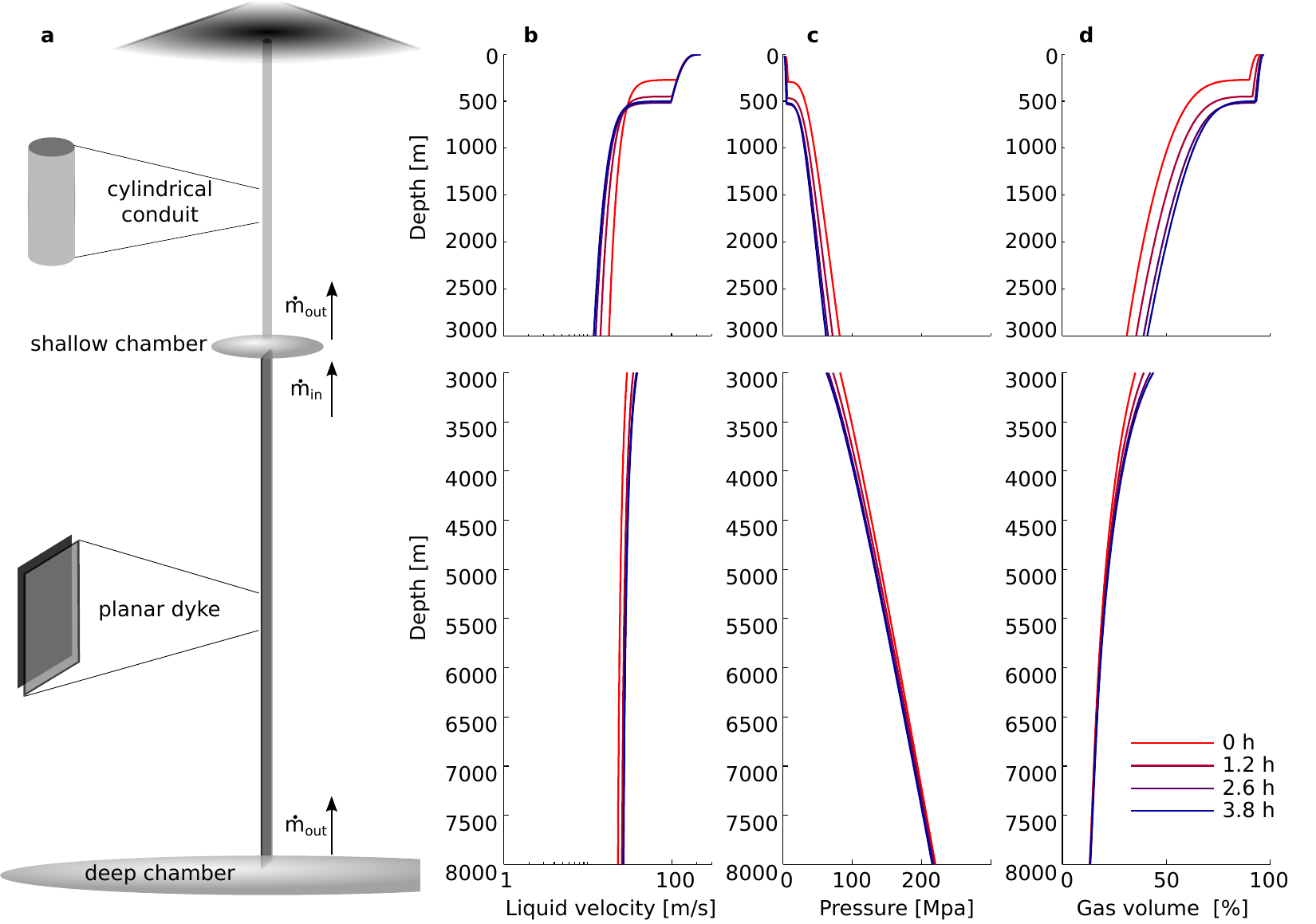} 
\caption{Simulation setup. a) Simulation domain. b-d) Example calculations at different times. Range of employed input parameters: shallow and deep chamber volumes 1.5 -- 30 and 200 -- 20,000 $\mathrm{km^3}$, respectively; shallow and deep chamber depths 2 -- 5 and 8 km, respectively; conduit diameter 50 -- 120 m; dyke width 4 -- 30 m; total water and carbon dioxide in the shallow dacite 2 -- 6 and 0 -- 1 wt\%, respectively; total water and carbon dioxide in the deep andesite 3 -- 7 and 0 -- 2 wt\%, respectively. Andesitic and dacitic magma compositions are reported in the Supplementary Table S1.}
\end{figure}

\begin{figure}
\center
\includegraphics[width=1\linewidth]{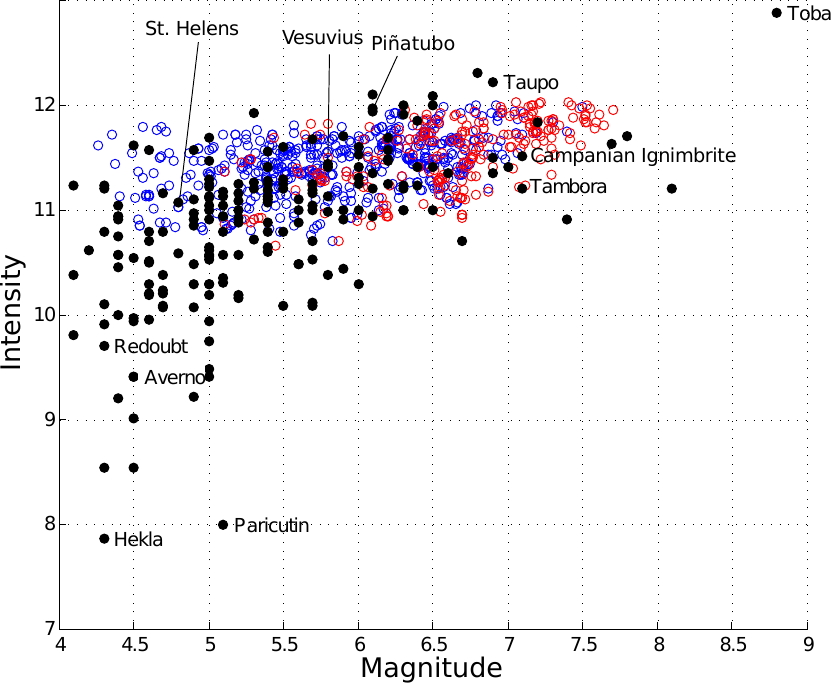}
\caption{Observed (filled circles) vs. simulated (blue and red empty circles) magnitude-intensity distribution. Data from the LaMEVE database (\cite{crosweller2012,brown2014}; www.bgs.ac.uk/vogripa, accessed in February 2019). Magnitude and intensity as in \cite{pyle2015}, also defined in the Methods. The space of variables has been sampled through a Latin Hypercube Sampling (LHS) technique \cite{adams2014}. Among the simulated points, blue refers to cases for which eruption closure is due to shallow chamber collapse, and red to cases for which eruption closure is due to deep chamber collapse.}
\end{figure}

\begin{figure}
\center
\includegraphics[width=1\linewidth]{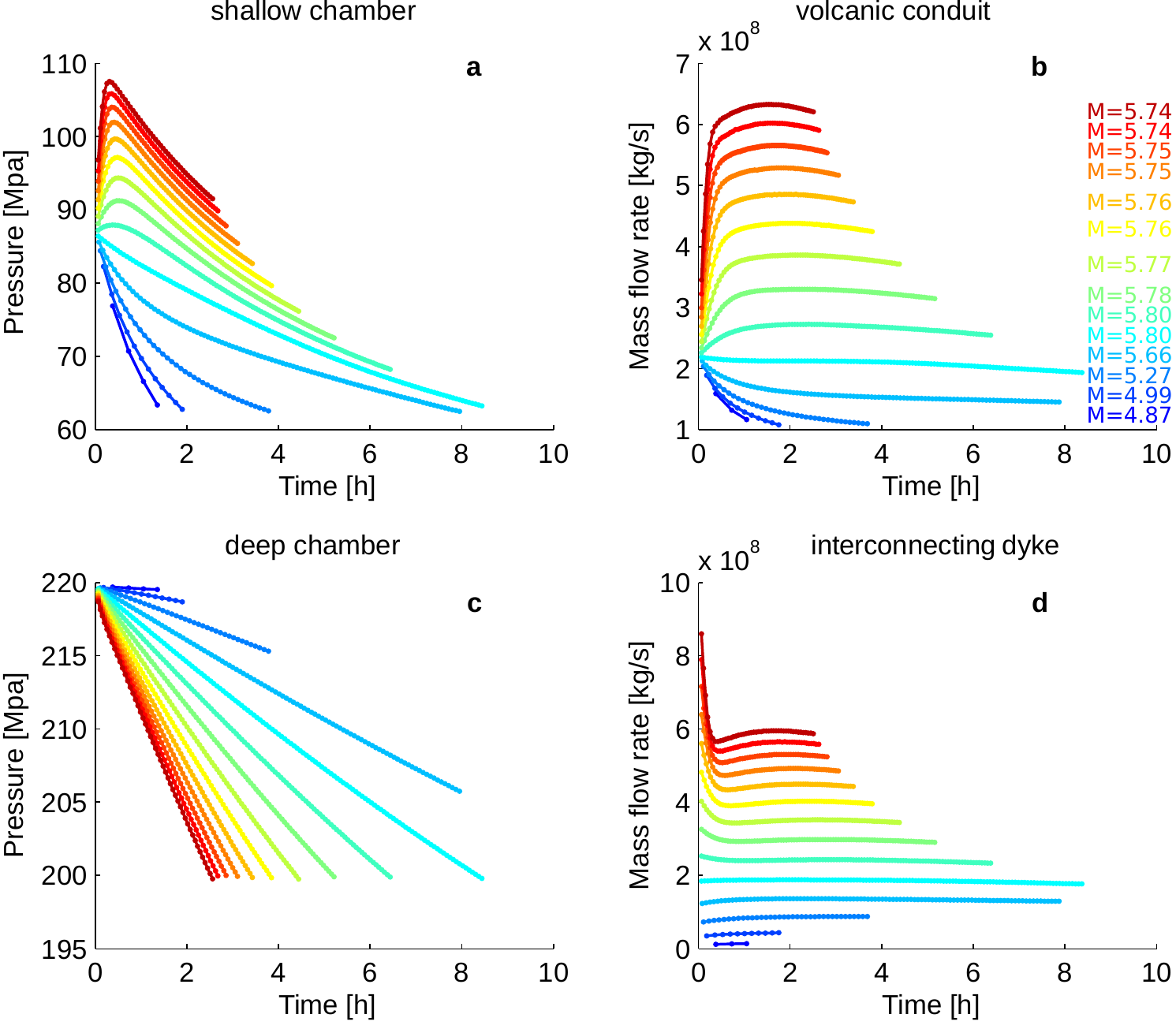}  
\caption{Computed pressure and mass flow rate. Calculations corresponding to dyke widths from 4 (cold colours) to 30 (hot colours) m, with regular dyke width increment by 2 m. Simulation conditions: deep chamber volume = 200 $\mathrm{km^3}$; shallow chamber volume = 3 $\mathrm{km^3}$; depth of the shallow chamber = 3 km; conduit diameter = 90 m; total volatile contents: H2O in dacite = 4.5 wt\%, CO2 in dacite = 1 w\%, H2O in andesite = 5 wt\%, CO2 in andesite = 2 wt\%. The quantity M in panel 3b is the computed eruption magnitude. The corresponding evolutions of magma density, gas volume fraction, and erupted magma composition, are reported in the Supplementary Figures S1-S2.}
\end{figure}

\begin{figure}
\center
\includegraphics[width=0.5\linewidth]{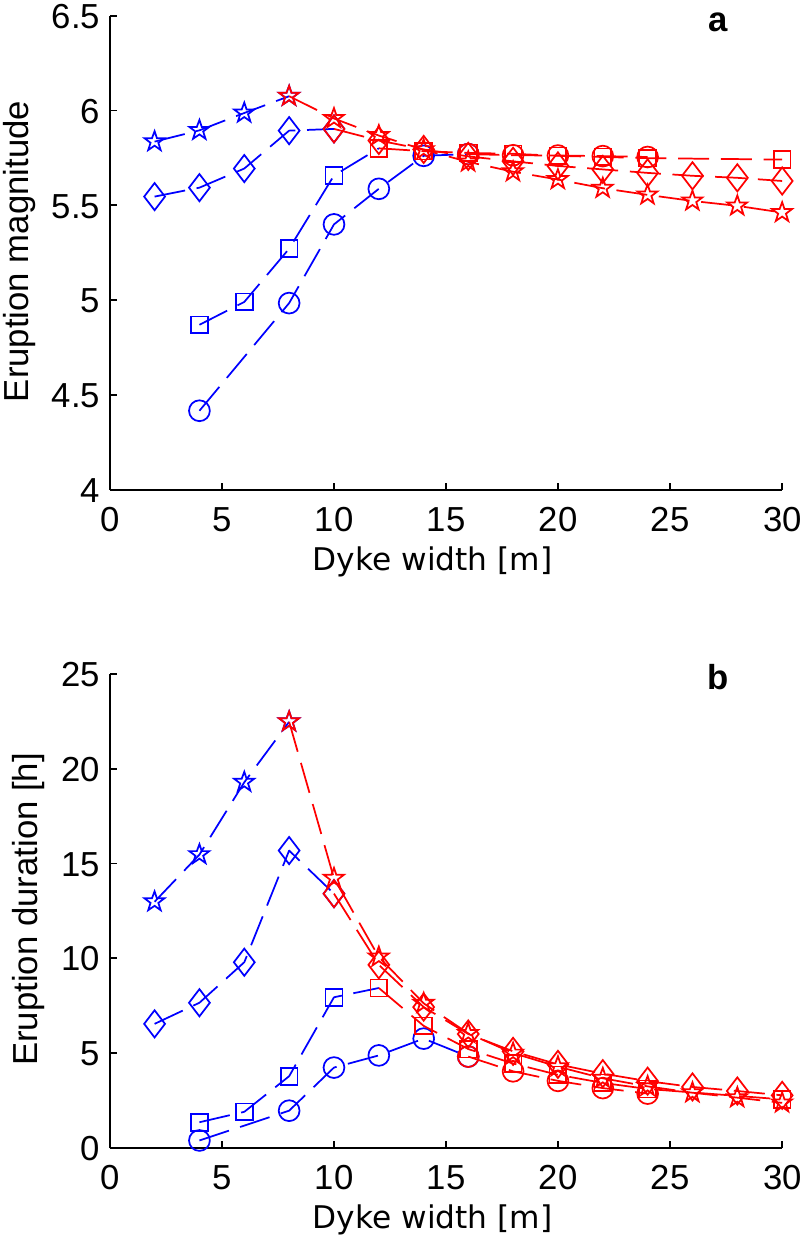} 
\caption{Role of deep interconnections. Calculated eruption magnitude (a) and duration (b) as a function of dyke width, for deep chamber volume = 200 $\mathrm{km^3}$ and other conditions as in Fig. 3. The different symbols refer to different shallow chamber volumes: circles = 1.5 $\mathrm{km^3}$; squares = 3 $\mathrm{km^3}$; diamonds = 15 $\mathrm{km^3}$; stars = 30 $\mathrm{km^3}$. Symbol colours as in Fig. 2: blue and red indicate eruption closure by shallow or deep chamber collapse, respectively.}
\end{figure}

\clearpage

\bibliography{biblio}

\begin{thebibliography}{10}

\bibitem{adams2014}
B.~Adams et~al.
\newblock Dakota, a multilevel parallel object-oriented framework for design
  optimization, parameter estimation, uncertainty quantification, and
  sensitivity analysis: Version 6.0 user’s manual.
\newblock {\em Sandia Technical Report SAND2014-4633 (2014).}

\bibitem{araya2019}
N.~Araya, M.~Nakamura, A.~Yasuda, S.~Okumura, T.~Sato, M.~Iguchi, D.~Miki, and
  N.~Geshi.
\newblock Shallow magma pre-charge during repeated plinian eruptions at
  sakurajima volcano.
\newblock {\em Scientific reports}, 9:1979, 2019.

\bibitem{bower1998}
S.~M. Bower and A.~Woods.
\newblock On the influence of magma chambers in controlling the evolution of
  explosive volcanic eruptions.
\newblock {\em Journal of Volcanology and Geothermal Research}, 86:67--78,
  1998.

\bibitem{brown2014}
S.~Brown, H.~Crosweller, R.~Sparks, et~al.
\newblock Global database on large magnitude explosive volcanic eruptions
  (lameve).
\newblock {\em J Appl. Volcanol.}, 5(3), 2014.

\bibitem{colucci2014}
S.~Colucci, M.~{de' Michieli Vitturi}, A.~Neri, and D.~Palladino.
\newblock An integrated model of magma chamber, conduit and column for the
  analysis of sustained explosive eruptions.
\newblock {\em Earth and Planetary Science Letters}, 404:98 -- 110, 2014.

\bibitem{colucci2017}
S.~Colucci, P.~Papale, and C.~P. Montagna.
\newblock Non-newtonian flow of bubbly magma in volcanic conduits.
\newblock {\em Journal of Geophysical Research: Solid Earth},
  122(3):1789--1804, 2017,
  https://agupubs.onlinelibrary.wiley.com/doi/pdf/10.1002/2016JB013383.

\bibitem{crosweller2012}
H.~Crosweller, B.~Arora, S.~Brown, et~al.
\newblock Global database on large magnitude explosive volcanic eruptions
  (lameve).
\newblock {\em J Appl. Volcanol.}, 4(1), 2012.

\bibitem{edmonds2019}
M.~Edmonds, K.~V. Cashman, M.~Holness, and M.~Jackson.
\newblock Architecture and dynamics of magma reservoirs.
\newblock {\em Philosophical Transactions of the Royal Society A: Mathematical,
  Physical and Engineering Sciences}, 377(2139):20180298, 2019,
  https://royalsocietypublishing.org/doi/pdf/10.1098/rsta.2018.0298.

\bibitem{folch1998}
A.~Folch and J.~Marti.
\newblock The generation of overpressure in felsic magma chambers by
  replenishment.
\newblock {\em Earth and Planetary Science Letters}, 163(1):301 -- 314, 1998.

\bibitem{folch2009}
A.~Folch and J.~Marti.
\newblock Time-dependent chamber and vent conditions during explosive
  caldera-forming eruptions.
\newblock {\em Earth and Planetary Science Letters}, 280(1):246 -- 253, 2009.

\bibitem{Gudmundsson2007}
A.~Gudmundsson.
\newblock Conceptual and numerical models of ring-fault formation.
\newblock {\em Journal of Volcanology and Geothermal Research}, 164(3):142 --
  160, 2007.

\bibitem{gudnason2018}
J.~Gudnason, T.~Thordarson, B.~F. Houghton, and G.~Larsen.
\newblock The 1845 hekla eruption: Grain-size characteristics of a tephra
  layer.
\newblock {\em Journal of Volcanology and Geothermal Research}, 350:33 -- 46,
  2018.

\bibitem{Kabele2017}
P.~Kabele, J.~Žák, and M.~Somr.
\newblock {Finite-element modeling of magma chamber–host rock interactions
  prior to caldera collapse}.
\newblock {\em Geophysical Journal International}, 209(3):1851--1865, 03 2017,
  https://academic.oup.com/gji/article-pdf/209/3/1851/14012135/ggx121.pdf.

\bibitem{macedonio2005}
G.~Macedonio, A.~Neri, J.~Marti, and A.~Folch.
\newblock Temporal evolution of flow conditions in sustained magmatic explosive
  eruptions.
\newblock {\em Journal of Volcanology and Geothermal Research}, 143(1):153 --
  172, 2005.
\newblock Volcanic Eruption Mechanisms.

\bibitem{marti2000}
J.~Martı, A.~Folch, A.~Neri, and G.~Macedonio.
\newblock Pressure evolution during explosive caldera-forming eruptions.
\newblock {\em Earth and Planetary Science Letters}, 175(3):275 -- 287, 2000.

\bibitem{miller1994}
T.~Miller and B.~Chouet.
\newblock The 1989-1990 eruptions of redoubt volcano: an introduction.
\newblock {\em Journal of Volcanology and Geothermal Research}, 62(1):1 -- 10,
  1994.

\bibitem{neri1998}
A.~Neri, P.~Papale, and G.~Macedonio.
\newblock The role of magma composition and water content in explosive
  eruptions: 2. pyroclastic dispersion dynamics.
\newblock {\em Journal of Volcanology and Geothermal Research}, 87(1):95 --
  115, 1998.

\bibitem{papale2001}
P.~Papale.
\newblock Dynamics of magma flow in volcanic conduits with variable
  fragmentation efficiency and nonequilibrium pumice degassing.
\newblock {\em Journal of Geophysical Research: Solid Earth},
  106(B6):11043--11065, 2001,
  https://agupubs.onlinelibrary.wiley.com/doi/pdf/10.1029/2000JB900428.

\bibitem{papale2018}
P.~Papale.
\newblock Global time-size distribution of volcanic eruptions on earth.
\newblock {\em Scientific Reports}, 8:6838, 2018.

\bibitem{pioli2008}
L.~Pioli, E.~Erlund, E.~Johnson, K.~Cashman, P.~Wallace, M.~Rosi, and
  H.~{Delgado Granados}.
\newblock Explosive dynamics of violent strombolian eruptions: The eruption of
  parícutin volcano 1943–1952 (mexico).
\newblock {\em Earth and Planetary Science Letters}, 271(1):359 -- 368, 2008.

\bibitem{pistolesi2016}
M.~Pistolesi, R.~Isaia, P.~Marianelli, A.~Bertagnini, C.~Fourmentraux, P.~G.
  Albert, E.~L. Tomlinson, M.~A. Menzies, M.~Rosi, and A.~Sbrana.
\newblock {Simultaneous eruptions from multiple vents at Campi Flegrei (Italy)
  highlight new eruption processes at calderas}.
\newblock {\em Geology}, 44(6):487--490, 06 2016,
  https://pubs.geoscienceworld.org/geology/article-pdf/44/6/487/3550953/487.pdf.

\bibitem{pouget2013}
S.~Pouget, M.~Bursik, P.~Webley, J.~Dehn, and M.~Pavolonis.
\newblock Estimation of eruption source parameters from umbrella cloud or
  downwind plume growth rate.
\newblock {\em Journal of Volcanology and Geothermal Research}, 258:100 -- 112,
  2013.

\bibitem{pyle2015}
D.~M. Pyle.
\newblock Chapter 13 - sizes of volcanic eruptions.
\newblock In H.~Sigurdsson, editor, {\em The Encyclopedia of Volcanoes (Second
  Edition)}, pages 257 -- 264. Academic Press, Amsterdam, second edition
  edition, 2015.

\end{thebibliography}


\begin{thebibliography}{10}

\bibitem{aravena2017}
A.~Aravena, M.~{de' Michieli Vitturi}, R.~Cioni, and A.~Neri.
\newblock Stability of volcanic conduits during explosive eruptions.
\newblock {\em Journal of Volcanology and Geothermal Research}, 339:52 -- 62,
  2017.

\bibitem{bower1998}
S.~M. Bower and A.~Woods.
\newblock On the influence of magma chambers in controlling the evolution of
  explosive volcanic eruptions.
\newblock {\em Journal of Volcanology and Geothermal Research}, 86:67--78,
  1998.

\bibitem{colucci2014}
S.~Colucci, M.~{de' Michieli Vitturi}, A.~Neri, and D.~Palladino.
\newblock An integrated model of magma chamber, conduit and column for the
  analysis of sustained explosive eruptions.
\newblock {\em Earth and Planetary Science Letters}, 404:98 -- 110, 2014.

\bibitem{Gudmundsson2007}
A.~Gudmundsson.
\newblock Conceptual and numerical models of ring-fault formation.
\newblock {\em Journal of Volcanology and Geothermal Research}, 164(3):142 --
  160, 2007.

\bibitem{Kabele2017}
P.~Kabele, J.~Žák, and M.~Somr.
\newblock {Finite-element modeling of magma chamber–host rock interactions
  prior to caldera collapse}.
\newblock {\em Geophysical Journal International}, 209(3):1851--1865, 03 2017,
  https://academic.oup.com/gji/article-pdf/209/3/1851/14012135/ggx121.pdf.

\bibitem{macedonio1994}
G.~Macedonio, F.~Dobran, and A.~Neri.
\newblock Erosion processes in volcanic conduits and application to the ad 79
  eruption of vesuvius.
\newblock {\em Earth and Planetary Science Letters}, 121(1):137 -- 152, 1994.

\bibitem{macedonio2005}
G.~Macedonio, A.~Neri, J.~Marti, and A.~Folch.
\newblock Temporal evolution of flow conditions in sustained magmatic explosive
  eruptions.
\newblock {\em Journal of Volcanology and Geothermal Research}, 143(1):153 --
  172, 2005.
\newblock Volcanic Eruption Mechanisms.

\bibitem{papale2001}
P.~Papale.
\newblock Dynamics of magma flow in volcanic conduits with variable
  fragmentation efficiency and nonequilibrium pumice degassing.
\newblock {\em Journal of Geophysical Research: Solid Earth},
  106(B6):11043--11065, 2001,
  https://agupubs.onlinelibrary.wiley.com/doi/pdf/10.1029/2000JB900428.

\bibitem{papale1994}
P.~Papale and F.~Dobran.
\newblock Magma flow along the volcanic conduit during the plinian and
  pyroclastic flow phases of the may 18, 1980, mount st. helens eruption.
\newblock {\em Journal of Geophysical Research: Solid Earth},
  99(B3):4355--4373, 1994,
  https://agupubs.onlinelibrary.wiley.com/doi/pdf/10.1029/93JB02972.

\bibitem{papale2006}
P.~Papale, R.~Moretti, and D.~Barbato.
\newblock The compositional dependence of the saturation surface of h2o+co2
  fluids in silicate melts.
\newblock {\em Chemical Geology}, 229(1):78 -- 95, 2006.
\newblock Physics, Chemistry and Rheology of Silicate Melts and Glasses.

\bibitem{pyle2015}
D.~M. Pyle.
\newblock Chapter 13 - sizes of volcanic eruptions.
\newblock In H.~Sigurdsson, editor, {\em The Encyclopedia of Volcanoes (Second
  Edition)}, pages 257 -- 264. Academic Press, Amsterdam, second edition
  edition, 2015.

\end{thebibliography}
\bibliographystyle{habbrv}

\end{document}


\maketitle

\section{Methods}

Our computations refer to the sub-steady eruption phases for which the time scale of system evolution is significantly longer than the transit time of magma in the simulated domain. Boundary conditions for dyke and conduit flow reflect magma withdrawal (deep chamber) and balance between magma input and output under instantaneous magma mixing (shallow chamber). The upflow conditions for dyke flow correspond to deep chamber conditions. Pressure in the shallow chamber represents both the downflow boundary condition for dyke flow, and the upflow boundary condition for conduit flow, and together with the evolution of magma composition upon mixing in the shallow chamber, it couples dyke and conduit flow dynamics, and ultimately results in consistent evolution of the entire simulated system from the deeper chamber to volcanic conduit.

 The fundamental equations describing transport of mass and momentum along a conduit or dyke under 1D, steady, isothermal, multiphase conditions are reported in \cite{papale2001}.
A lumped system approximation for the two magma chambers keeps the overall complexity to manageable levels, at the same time eliminating further arbitrary variables and allowing the extraction of first order controls on the eruption dynamics by the composite nature of the plumbing system. The computation is terminated when the magmatic pressure somewhere in the simulated domain falls below the local lithostatic pressure by a quantity exceeding the rock tensile strength, here fixed to an average value of 20 MPa \cite{bower1998,Gudmundsson2007,Kabele2017}. In details:
\begin{enumerate}

 \item  At each time step, the equations for the cylindrical upper conduit and for the planar dyke connecting the deep reservoir to the shallow chamber are solved, with boundary conditions given by pressure and composition in the two chambers.

\item At time zero, we set a lithostatic pressure boundary condition for both chambers. Initial pressurized conditions are also possible, but they are not considered here in order to keep to a minimum the number of independent variables characterising the system. The conduit/dyke exit boundary conditions are determined by either choked flow or ambient pressure (atmopheric pressure for the conduit, and shallow chamber pressure for the dyke), as part of the solution of the simulated dynamics. Note that time zero does not correspond to the start of the eruption, which would be characterised by highly transient dynamics during which the volcanic conduit and eruption plume are established. These transient dynamics are not considered in our modelling. Accordingly, time zero corresponds to a reference time when steady flow conditions have been established along the entire simulated domain from the deep chamber to the surface. 

\item For later times, composition and pressure inside each one of the two chambers is computed on the basis of mass inflow/outflow, and constitute the new boundary condition for the numerical calculations of dyke and conduit flow. For dyke flow, the third, not simulated dimension has been fixed to 100 m when converting from mass flow rate per unit length (computed through dyke ascent modelling) to mass flow rate, and from that, to chamber mass loss or gain (see below). In particular, after update of mixture density given by
 \begin{equation}\label{eq1}
 \rho_m^{t+\Delta t} = \frac{m^{t+\Delta t}}{v^{t+\Delta t}}, 
 \end{equation}
 where $\rho_m$ is mixture density, $t$ is time, $m$ is mass inside the chamber, $v$ is chamber volume, the corresponding pressure is calculated. In fact, at thermodynamic equilibrium assumed in the simulations, mixture density and pressure are univocally related via the real equations of state for the liquid and gas phases and the multi-component gas-melt equilibrium model employed \cite{papale2006}. Such a relationship is non-linear, requiring a numerical solution for pressure.  That implies determination of the multi-component volatile saturation surface and melt/gas densities at chamber conditions. 
 
  Under the quasi-steady flow assumption, the mass inside the chamber is approximated as
\begin{equation}\label{eq2}
 m^{t+\Delta t} \approx m^t + \left(\dot{m}_{in}^{t}-\dot{m}_{out}^{t}\right)\Delta t,
\end{equation}
where $\dot{m}$ is mass flow rate. Chamber volume is approximated by
\begin{equation}\label{eq3}
 v^{t+\Delta t} \approx v^t + \frac{v^t \Delta P^t}{\beta},
\end{equation}
where $\beta$ is the effective bulk modulus of the elastic walls surrounding the chambers. By repeating the same simulations with $\beta = \mathrm{10^{10}}$ Pa \cite{bower1998} and $\beta=\infty$ (non-deformable chamber walls), we have found that inclusion of rock elasticity and associated chamber volume changes produce minor or negligible differences in the simulated dynamics (see the Supplementary Figure S3). Therefore, we refer to only the latter case ($\beta=\infty$, non-deformable chamber walls) throughout this paper.

\item  The time step $\Delta t$ used to advance along the sequence of steady states is determined at run time as twice the longest among the transit times in the conduit and dyke
\begin{equation}
\Delta t = 2*t_{trans}.
\end{equation}
The transit time $t_{trans}$ into the conduit/dyke is computed by space integration of the inverse of velocity.

\item The cycle stops when the local difference between the lithostatic and magmatic pressure anywhere in the computational domain exceeds the tensile strenght of rocks. 

\item A-posteriori evaluation of the quasi-steady assumption is done by checking if the following criterion is satisfied:
\begin{equation}
t_{\dot{m}} >> t_{trans},
\end{equation}
where $t_{\dot{m}}$ is the characteristic time of variation of the mass flow rate over $\Delta t$
\begin{equation}
 t_{\dot{m}} = \frac{\dot{m}^{t+\Delta t}}{|\dot{m}^{t+\Delta t}-\dot{m}^{t}|}\Delta t.
\end{equation}

\end{enumerate}

In spite of its relative simplicity, such a set up captures the fundamental characteristic of many volcanic plumbing systems characterised by separated reservoirs located at different depth and hosting different magmas, being simultaneously activated during an eruption. Comparison with real eruptions mostly involves the quantities Magnitude and Intensity (see Fig. 2). Magnitude (M) is computed as in \cite{pyle2015} as $M=\log[{m}_T (kg)]-7$, where $m_T$ is the total mass erupted; in the simulations, $m_T$ is determined by integrating in time the computed mass flow rate, see panel b in Fig. 3. Intensity (I) refers to the eruption mass flow rate, and is defined as in \cite{pyle2015} as $I=\log[\dot{m}(kg/s)]+3$, where $\dot{m}$ is the average mass flow rate.

Point 5 above requires further discussion. As explained above, we stop a simulation run when the pressure somewhere in the simulated domain, extending from the deep chamber to the surface, falls below the critical threshold dictated by rock tensile strength. Beyond such conditions local rock collapse is expected to occur, terminating the sub-steady phase of the eruption. When such a condition is found in correspondence of one of the magma chambers, a caldera collapse may occur. In the real world the processes may be more complex: in some cases the initiation of caldera roof collapse may cause the magmatic pressure to increase back to values sufficient to avoid further collapse, and a sub-steady phase may be restored until further magmatic pressure changes lead to new instabilities. Similarly, if large pressure decrease occurs locally in the volcanic conduit (typically close to magma fragmentation where the pressure gradient is the highest \cite{papale1994}), that may lead to local conduit wall erosion and conduit shape changes \cite{macedonio1994,aravena2017} and to transient eruption dynamics. Additional complexities relate to the mechanics of rock failure \cite{Kabele2017}, especially in layered environments \cite{Gudmundsson2007}, whereby the generation of ring faults and the occurrence of caldera collapse depend on a number of other factors including chamber shape. Because we only refer to steady flow dynamics and do not solve the rock mechanics associated to fracturing and faulting, that are beyond the scopes of this work, we stop our simulations when the computed pressure changes do not guarantee further validity of the steady flow assumption. Such an approach is similar to that employed by previous authors \cite{bower1998,macedonio2005,colucci2014} who considered magma withdrawal from one single chamber. 

The CONDUIT4 code \cite{papale2001} used for the simulations of magma flow along the conduit and dyke is available upon request to the authors. The SOLWCAD code \cite{papale2006} for multi-component volatile saturation is embedded in the CONDUIT4 code, and openly available from \url{www.pi.ingv.it/progetti/eurovolc}.

\clearpage

\renewcommand{\thetable}{S1}

\begin{table}[t]
  \centering
  \caption{Chemical composition of the volatile-free liquid magmas employed in the simulations. }
  \vspace{4mm}
\begin{tabular}{|c|c|c|c|c|c|c|c|c|c|c|}
\hline
{} & {$\mathrm{SiO_2}$} & {$\mathrm{TiO_2}$} & {$\mathrm{Al_2O_3}$}& {$\mathrm{Fe_2O_3}$}& {$\mathrm{FeO}$} & {$\mathrm{MnO}$} & {$\mathrm{MgO}$} & {$\mathrm{CaO}$} & {$\mathrm{Na_2O}$} & {$\mathrm{K_2O}$}\\
\hline 
\hline
{Andesite} & {$\mathrm{59.85}$} & {$\mathrm{0.58}$} & {$\mathrm{ 18.30}$}& {$\mathrm{0}$}& {$\mathrm{6.62}$} & {$\mathrm{0.13}$} & {$\mathrm{2.81}$} & {$\mathrm{7.45}$} & {$\mathrm{ 3.34}$} & {$\mathrm{ 0.92}$}\\
\hline
{Dacite} & {$\mathrm{ 65.83}$} & {$\mathrm{ 0.65}$} & {$\mathrm{16.93}$}& {$\mathrm{4.02}$}& {$\mathrm{0}$} & {$\mathrm{0.07}$} & {$\mathrm{1.83}$} & {$\mathrm{4.24}$} & {$\mathrm{ 4.42}$} & {$\mathrm{ 2.00}$}\\
\hline
\end{tabular}

 \label{tab2}
\end{table}

\clearpage

\renewcommand{\thefigure}{S1}

\begin{figure}[t]
\centering
\includegraphics{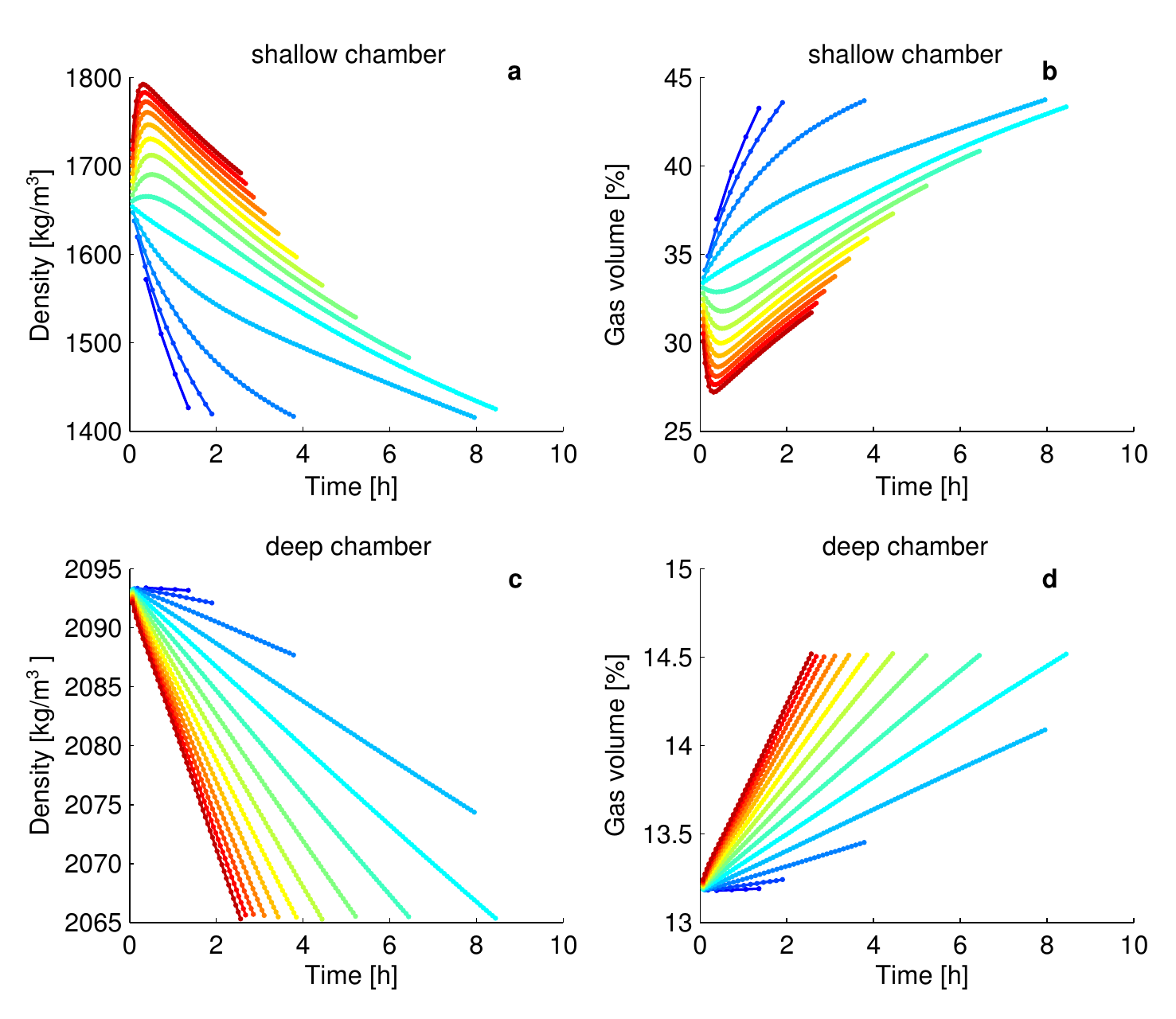}
\caption{Time series for density and gas volume. Calculated evolution of magma density (a and c, referred to the gas-melt mixture) and gas volume (b and d) inside the shallow and deep chamber, for the same simulation cases as in Fig. 3. }
\label{figS1}
\end{figure}

\renewcommand{\thefigure}{S2}

\begin{figure}[t]
\centering
\includegraphics{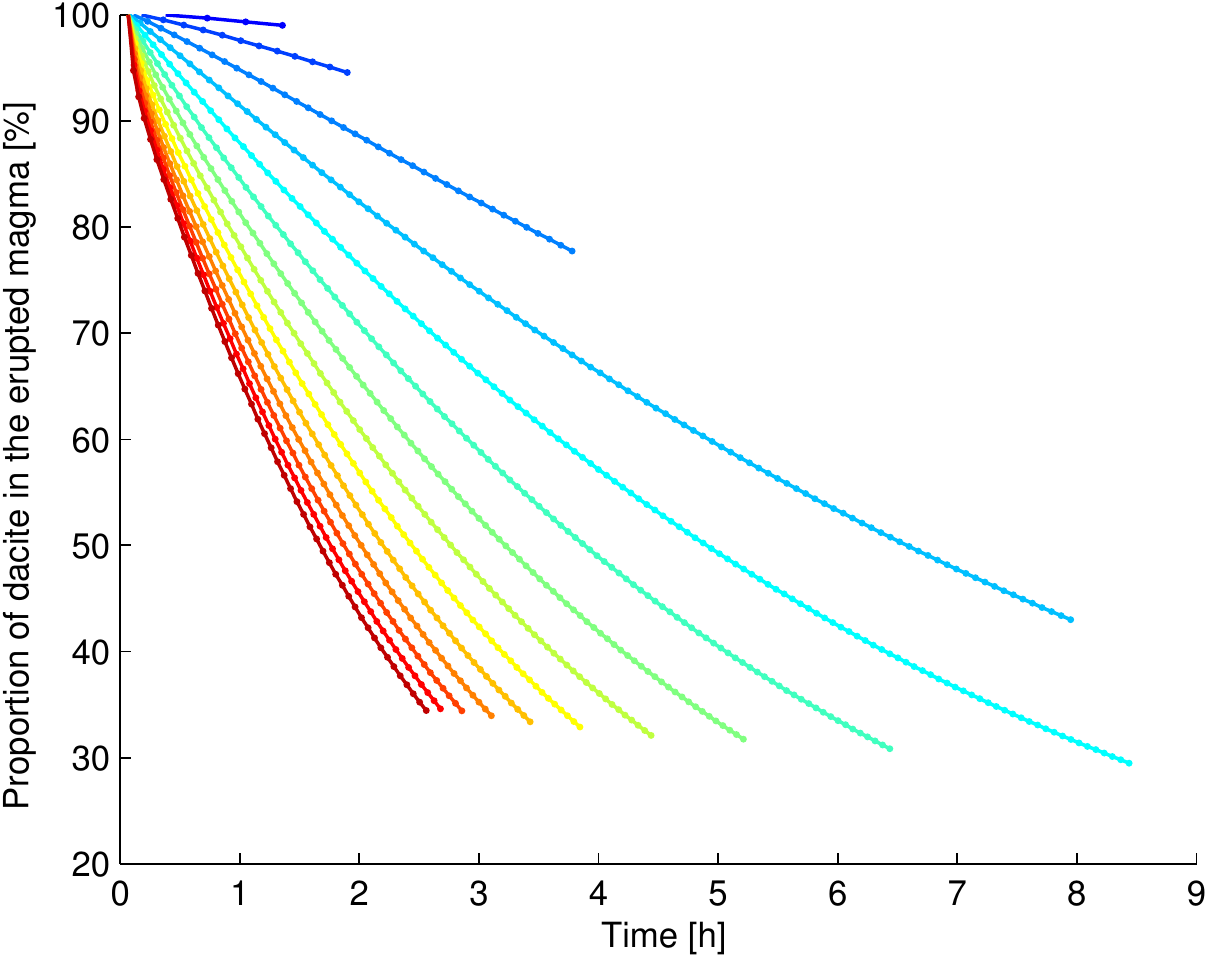}
\caption{Calculated evolution of the average composition of the erupted magma, for the same simulation cases as in Figs. 3 and S1.}
\label{figS2}
\end{figure}

\renewcommand{\thefigure}{S3}

\begin{figure}[t]
\centering
\includegraphics{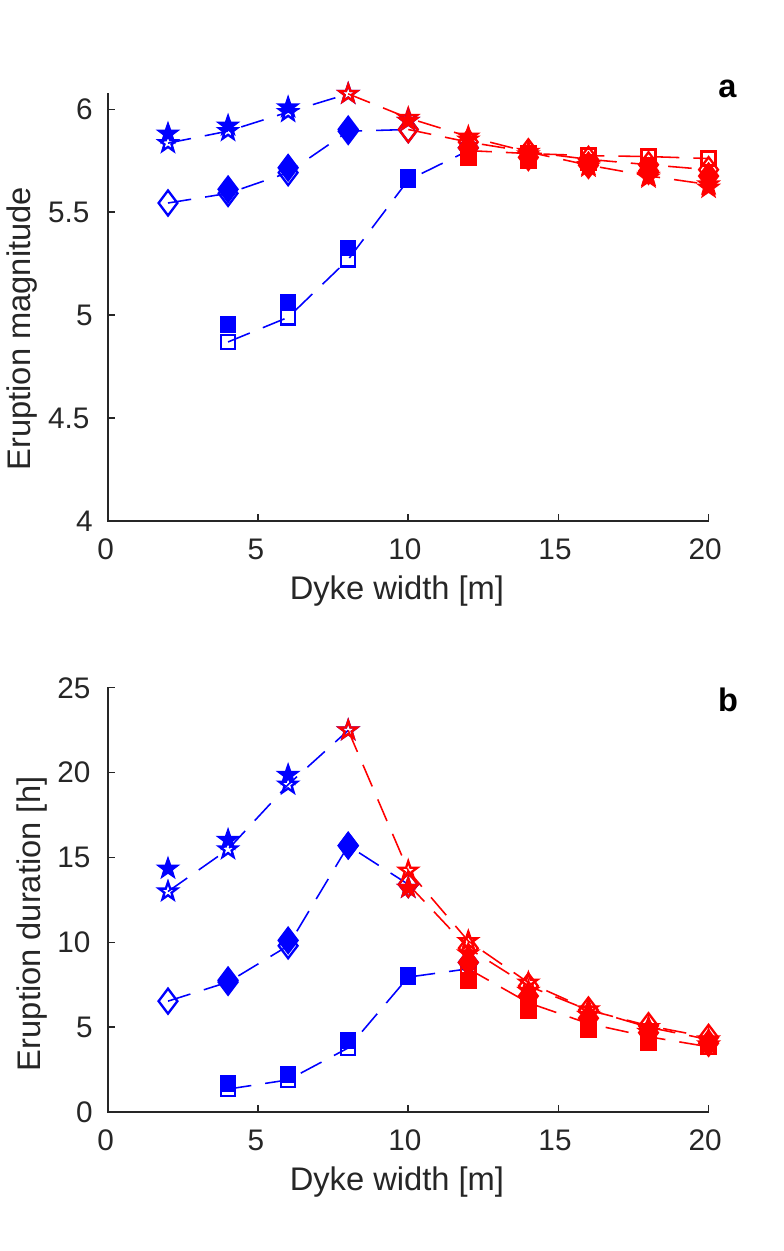}
\caption{The effects of chamber wall elasticity on computed eruption magnitude and duration. Open symbols as in Fig. 4 correspond to assuming rigid walls. Solid symbols assume instead elastic walls with bulk modulus equal to $10^{10}$ Pa. Note that most of the solid symbols are superimposed to the open ones.}
\label{figS3}
\end{figure}

\clearpage

\bibliography{biblio}
\bibliographystyle{habbrv}